\documentclass[conference]{IEEEtran}
\makeatletter
\def\footnoterule{\relax%
  \kern-5pt
  \hbox to \columnwidth{\hfill\vrule width 0.8\columnwidth height 0.4pt\hfill}
  \kern4.6pt}
\makeatother
\usepackage{graphicx}
\usepackage{cite}
\graphicspath{ {}}
\usepackage{balance}
\usepackage{enumitem}
\usepackage{amsmath}
\usepackage{amssymb}
\usepackage{listings}
\usepackage{subcaption}
\usepackage{comment}
\usepackage[none]{hyphenat}
\usepackage[dvipsnames]{xcolor}
\usepackage{amsmath}
\usepackage{url}
\usepackage{booktabs}
\usepackage{multirow}
\usepackage{hyperref}
\hypersetup{
    colorlinks=true,
    linkcolor=black,
    filecolor=ForestGreen,      
    urlcolor=blue,
    citecolor = black
}

\ifCLASSINFOpdf
\else
\fi
\begin{document}

\title{In-Situ Thickness Measurement of Die Silicon Using Voltage Imaging for Hardware Assurance}

\author{
\IEEEauthorblockN{
Olivia P. Dizon-Paradis\IEEEauthorrefmark{1},
Nitin Varshney\IEEEauthorrefmark{1},
M Tanjidur Rahman\IEEEauthorrefmark{1},\\ 
Michael Strizich\IEEEauthorrefmark{2},
Haoting Shen\IEEEauthorrefmark{3}, and
Navid Asadizanjani\IEEEauthorrefmark{1}\\}

\IEEEauthorblockA{\IEEEauthorrefmark{1}
Electrical and Computer Engineering Department, University of Florida\\
Gainesville, FL, USA\\
Email: nasadi@ece.ufl.edu\\}

\IEEEauthorblockA{\IEEEauthorrefmark{2}
MicroNet Solutions Inc.\\
Albuquerque, NM, USA \\
Email: mstriz@micronetsol.net}

\IEEEauthorblockA{\IEEEauthorrefmark{3}
Computer Science and Engineering Department, University of Nevada\\
Reno, NV, USA\\
Email: hshen@unr.edu}
}

\maketitle

\begin{abstract} 
Hardware assurance of electronics is a challenging task and is of great interest to the government and the electronics industry. 
Physical inspection-based methods such as reverse engineering (RE) and Trojan scanning (TS) play an important role in hardware assurance. 
Therefore, there is a growing demand for automation in RE and TS. 
Many state-of-the-art physical inspection methods incorporate an iterative imaging and delayering workflow. In practice, uniform delayering can be challenging if the thickness of the initial layer of material is non-uniform. 
Moreover, this non-uniformity can reoccur at any stage during delayering and must be corrected. 
Therefore, it is critical to evaluate the thickness of the layers to be removed in a real-time fashion. Our proposed method uses electron beam voltage imaging, image processing, and Monte Carlo simulation to measure the  thickness of remaining silicon to guide a uniform delayering process\footnote{DISTRIBUTION STATEMENT A. Approved for public release: distribution is
unlimited.}.

\end{abstract}


\IEEEpeerreviewmaketitle

\section{Introduction}
Physical inspection for hardware assurance has gained a lot of attention over the past few years. There are still many challenges left open to address before one can have a fast and fully-automated system for physical inspection. Reverse  engineering (RE) \cite{RE_survey} and Trojan scanner (TS) \cite{vashistha2018trojan} are examples of the techniques used for physical inspection of integrated circuits (ICs). RE can fully analyze  an  IC and extract the structure, connectivity, and functionality of the chip, but is time-consuming and labor-intensive. TS can authenticate an IC much faster, but it does not provide information on the full functionality of the chip. 

RE is used for both honest and dishonest purposes. Intellectual property (IP) piracy, malicious modification, and hardware Trojan insertion are some examples for dishonest purposes. 
Honest applications include authenticating the intellectual property (IP), evaluation of archaic legacy parts, and product development. 

The typical IC RE workflow involves the four major steps \cite{RE_survey}: 1) sample preparation; 2) iterative imaging-delayering cycle; 3) annotation; and 4) netlist extraction. Sample preparation exposes the silicon die of the chip. This step includes IC package removal by mechanical polishing and chemical etching, planarization, and cleaning \cite{RE_survey}. 

Once the die is exposed, iterative delayering (also known as deprocessing) and imaging evaluates the chip layer-by-layer at the metal and transistor level. 
Methods such as plasma/chemical etching and mechanical polishing are used to remove each layer of the IC to prepare the individual layers for imaging with optical or scanning electron microscopy (SEM). 

Delayering can be performed from the frontside (i.e. top metal layer of the chip) or the backside (i.e. substrate of the chip). 
Predominant industrial trends toward chips with more metal layers and higher densities of transistors with smaller technology nodes have made the backside more attractive for deprocessing \cite{plasma_FIB}.
In practice, inherent nonuniformity of the substrate recurringly raises a challenge in accurately estimating thickness during deprocessing. 
When the IC leaves the foundry, the silicon die is not perfectly level.
This nonuniformity is further exacerbated during RE due to complex thermal and mechanical deformations \cite{principe2017steps}. 

In TS, the image of the transistor layer, taken from the backside of the die, is compared with a golden sample/layout to detect any malicious modification. 
During the delayering, the bare die is first exposed by de-packaging the chip. However, such de-packaging exposes only the first layer of material, the silicon. 
To access the transistor layer of the circuit, further thinning is required by ultrathinning machines such as VarioMill by
Varioscale \cite{plasma_FIB}, which mechanically thins the die down to about 0.5 $\mu$m to 1.5 $\mu$m. 
During ultrathinning, it is critical to remove the remaining silicon as evenly as possible to preserve the transistor structure, as SEM imaging is very sensitive to the thickness of surface materials.
Therefore, similar to RE, nonunifomity of the die is also a challenge in TS. 

\begin{figure}
\centering
 \includegraphics[width=1\linewidth]{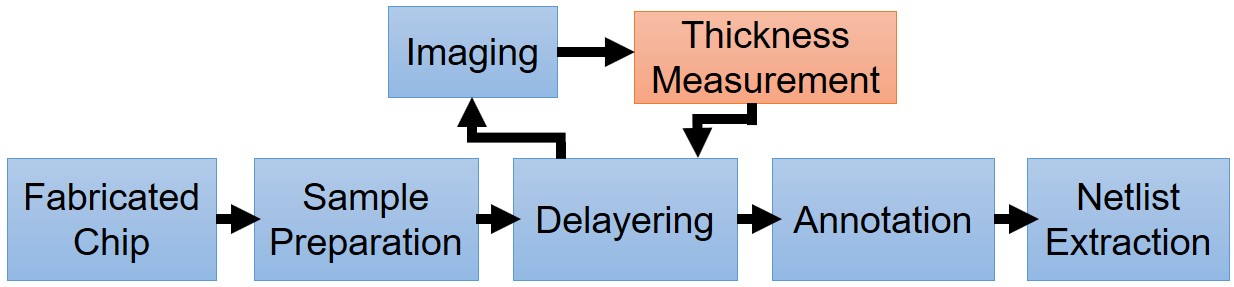}
\caption{Workflow for nanometer-scale reverse engineering with the proposed thickness measurement step.}
\label{fig: reverse engineering}
\end{figure}

In recent years, state-of-the-art SEM incorporated with focused ion beam (FIB) has made in-situ delayering a more practical approach for delayering of large areas. Therefore, a dynamically adaptable delayering procedure using FIB can be modified to address the challenge of uniform deprocessing. 
This would first require an evaluation of the thickness of the remaining silicon in the IC before adjusting the deprocessing rate at different locations of the die, as shown in Fig. \ref{fig: reverse engineering}. 
Thus, the main objective of this study is to develop a method to determine the real thickness of leftover silicon in an IC during the iterative imaging-delayering cycle. 
The proposed method uses electron beam (E-beam) voltage imaging, image processing, and Monte Carlo simulation to relate E-beam penetration depth and the real thickness of silicon for physical inspection.

\section{Methodology} \label{sec:methodology}

\begin{figure}[t]
\centering
\includegraphics[width=.65\linewidth]{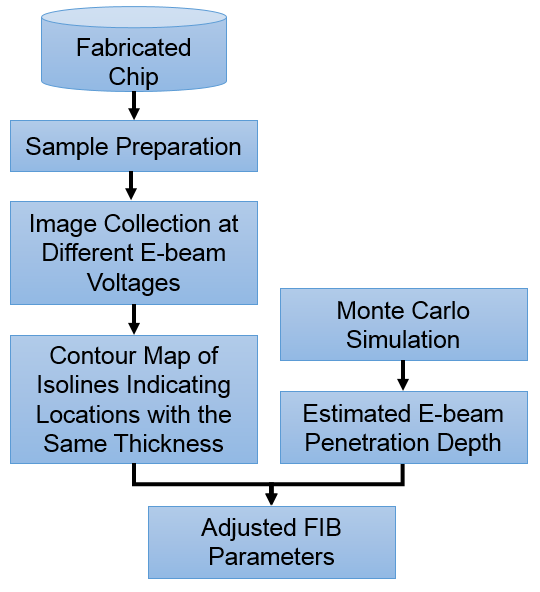} 
\caption{Workflow for thickness measurement.}
\label{fig: thickness measurement}
\end{figure}

In this section, we present a methodology for in-situ thickness measurement of left-over silicon substrate, summarized in Fig.~\ref{fig: thickness measurement}. In this methodology, an E-beam voltage source is used to obtain SEM images of the sample and image processing is used to generate a contour map of regions with similar thickness. In addition, a Monte-carlo simulation tool is used to calculate the penetration depth of the E-beam voltage. The calculated penetration depth is then used to estimate the amount of left-over silicon at the backside of the die. To validate this estimate, destructive testing is used to obtain the real thickness for comparison.

\subsection{Sample Preparation and Image Collection} \label{Sec:Sample_prep}

\begin{figure}
    \centering
    \includegraphics[width=0.75\linewidth]{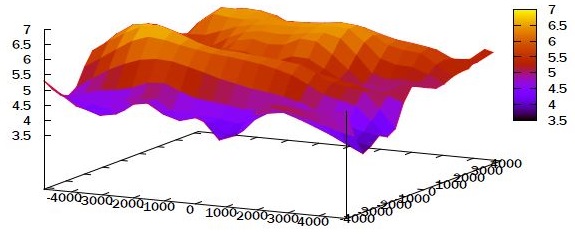}
    \caption{Real thickness map of an Intel SGX Prosessor after polishing to highlight the variation in the amount of silicon substrate left at different locations. All axis values are in $\mu$m.}
    \label{fig: Thickness_map}
\end{figure}

Though flip-chips are used in this study due to their widespread use in modern electronics, the proposed thickness measurement method is also applicable to non-flip chips. To prepare the samples for imaging, the outer package is first removed to expose the backside of the silicon die. For a non-flip chip, the packaging material can be removed by mechanical polishing or acid etching. For a flip-chip, the package can be removed by thermal and mechanical means\cite{rahmankey}. Once the silicon substrate is exposed, the bulk silicon is removed using selective polishing. The sample is then planarized and ultra-thinned from the backside of the chip with a precise 5-axis CNC mill. The surface topography of the ultra-thinned sample is nonuniform ~\cite{principe2017steps}. The thickness map of one such ultra-thinned sample showed a variety of thicknesses from 3~$\mu$m to 6.5~$\mu$m
(Fig.~\ref{fig: Thickness_map}). This demonstrates the surface nonuniformities that are introduced when backside ultra-thinning the entire chip. Hence, uniform backside delayering requires adjusting the FIB parameters according to the local thickness of the remaining silicon. 

Once the sample is ultra-thinned, the die is imaged with a TESCAN FERA3 dual beam microscope (FIB/SEM). The main objective of SEM imaging is to scan the whole die as fast as possible while still capturing sufficient feature details. This will significantly reduce the complexity and the processing time for the contour map generation step. SEM images can be collected using different electron beam (E-beam) voltages. The E-beam voltage, which denotes the beam energy, affects the penetration depth (i.e. the interaction volume of the electron beam within the sample) \cite{SEM_primer, hobler1995monte}. For example, images taken with lower electron voltages can expose active layers while images taken with higher electron voltages provide more detail from the sample's deeper subsurface features, including the polysilicon and higher metal layers. In addition, SEM images can be collected using two different modes: 1) back scattered electrons (BSE), and 2) secondary electrons (SE). In this study, BSE imaging is used because it offers better voltage contrast and fewer image artifacts.

\begin{figure}[t]
   \centering
   \includegraphics[width=.8\linewidth]{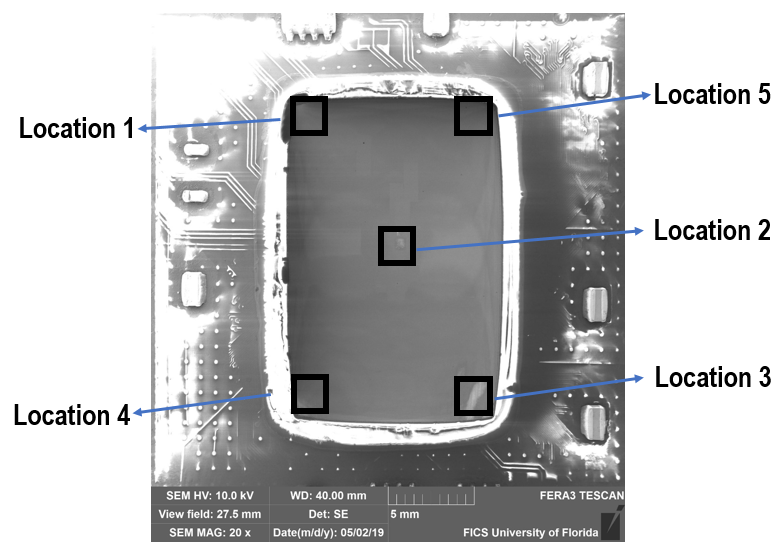} 
   \caption{Image collection locations on a Xilinx FPGA}
   \label{fig: image collect}
\end{figure}

We collected six BSE images at different E-beam voltages (15, 17, 21, 23, 25, and 30~kV) at five different locations on a Xilink FPGA (Fig. \ref{fig: image collect}). The collected images are saved as $1024\times1024$ pixel TIFF-formatted greyscales. In the following sections, Location 1 is focused on.

\subsection{Contour Map Generation using Image Processing} 
Since the thickness of the die is nonuniform, subsurface features appear in the collected images as the beam voltage varies. 
Image processing is then used to generate contour maps denoting the regions with similar amounts of remaining silicon. These contour maps are then used to adjust FIB parameters in a feedback loop to allow for automated uniform delayering. 

\begin{figure}
    \centering
    \includegraphics[width=1\linewidth]{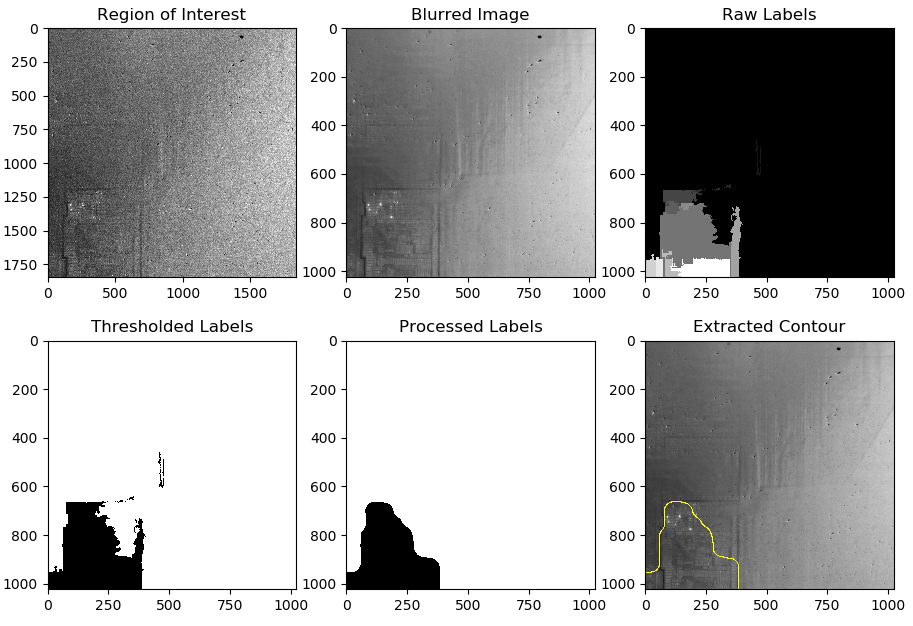}
    \caption{Image processing steps used to obtain contour maps, shown as applied on BSE images taken at Location 1 with 17kV.}
    \label{fig: image processing steps}
\end{figure}

The contouring steps are demonstrated in Fig. \ref{fig: image processing steps}. First, the image is cropped to the region of interest on the die and  blurred with a  Gaussian lowpass filter to isotopically reduce noise. Then, Felzenszwalb`s method is used to obtain a Superpixel over-segmentation \cite{felzenszwalb_efficient_2004}. Felzenszwalb`s method yields a higher number of small superpixels in regions with high contrast and a smaller number of large superpixels regions with low variation. Here, the regions where die structures are clearly visible have high contrast while regions where die structures are not visible have low contrast. Thus, we threshold the largest segments from the smallest segments based on size. The contours of the thresholded image are then noise-corrected and smoothed using morphological closing and opening. The resulting contour maps will be discussed in Section~\ref{sec: results_contour_map}.

\subsection{Estimated Thickness Measurement using Monte-Carlo Simulation}
Monte Carlo simulation is a well-established technique for estimating electron beam penetration, energy deposition, transmission, backscattering, and secondary emission, in both bulk matters and solids~\cite{monte_carlo}. Hence, there are several specialized programs available to simulate electron interactions in solids using Monte Carlo. In this study, we used CASINO and EISS to estimate the relationship between penetration depth and accelerating voltage~\cite{casino}\cite{eiss}. Many different test cases were explored by simulating electron trajectories at different accelerating voltages. In the following paragraphs, the mathematical theory used by these programs are detailed\cite{monte_carlo}\cite{math}.

First, the electron distribution is modeled as a Gaussian process from the origin of the beam. Then, Equation \ref{eq: gauss} below is used to determine which atom is responsible for the elastic scattering:

\begin{equation}
    Random > \sum^{n}_{i=1}\frac{\sigma_iF_i}{\sum^{n}_{j=1}\sigma_iF_i},
    \label{eq: gauss}
\end{equation}
 
where $Random$ is a random number uniformly distributed between 0 and 1, $\sigma_i$ is the total cross-section of element $i$, $F_i$ is the atomic fraction of element $i$, and $n$ is the number of elements in the region \cite{math}. When \ref{eq: gauss} is true, element i is responsible for the collision. The polar angle of collision, $\theta$, is computed by solving for $\theta$ in Equation \ref{eq: theta}:

\begin{equation}
    R = \frac{\int_0^\theta \frac{d\sigma}{d\theta}\sin{\theta}d\theta}{\int_0^\pi \frac{d\sigma}{d\theta}\sin{\theta}d\theta},
    \label{eq: theta}
\end{equation}

where $\frac{d\sigma}{d\theta}$ is the partial cross-section of element $i$ and R is another random number\cite{math}. The polar angle of collision is used to compute the distance, $L$, between two collisions. 
According to the continuous slowing down approximation (CDSA), the energy loss resulting from inelastic collisions between individual elastic scattering events can be simulated from stopping-power theory~\cite{monte_carlo}. Therefore, the energy lost during a travel distance, $L$, can be modeled as a constant value. Then, the energy at position $i$ along a trajectory is computed using the following equation:

\begin{equation}
    E_i = E_{i-1}+\frac{dE}{dS}L,
    \label{eq: energy}
\end{equation}
 
where $E_{i-1}$ and $E_i$ are the energies at the previous and current collision and $\frac{dE}{dS}$ is the rate of energy loss.
Both CASINO and EISS simulate the electron beam by scattering the electrons from one location, i.e. by single scattering. Then, the electron trajectories are simulated until they either exit the sample or lose all energy and come to rest within the sample.

\begin{figure}
    \centering
    \includegraphics[width=1\linewidth]{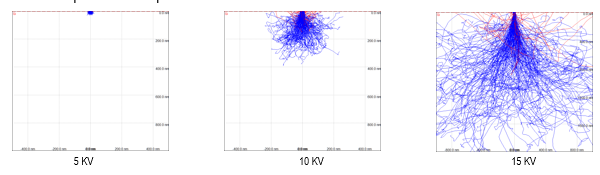}
    \caption{Monte Carlo simulation of electron trajectories in a bulk sample of silicon at different accelerating voltages to illustrate the direct correlation between accelerating voltage an interaction volume. Simulations were obtained using CASINO software~\cite{casino}. Trajectories of backscattered electrons are in red while trajectories of inelastically scattered electrons are in blue.}
    \label{fig: monte_carlo}
\end{figure}

From preliminary results shown in Fig. \ref{fig: monte_carlo}, it is clear that a reduction of the incident energy minimizes the interaction volume of the electron beam within the sample. The shape of this interaction volume does not change dramatically with energy, but its physical size does.
The red lines in the electron trajectories show the back-scattered electrons with enough energy to overcome the escape barrier of the surface. These electrons, which typically possess more energy than the secondary electrons, can be detected by back-scattered detectors and provide more information on deeper sub-surface features of the silicon substrate. This provides a rough estimate of the amount of  silicon left at different locations.

\subsection{Validation using a Commercial Instrument}
During delayering, the contour map generated from BSE images and the thickness measurement estimation from the Monte Carlo simulation are used to estimate the thickness of remaining silicon. To validate this estimated thickness, a commercial VarioMill instrument by Varioscale was used to obtain the real thickness map of remaining silicon for comparison.

\section{Results} \label{Sec:Result}

\subsection{Contour Map Generation using Image Processing} \label{sec: results_contour_map}
\begin{figure}
    \centering
    \includegraphics[width=1\linewidth]{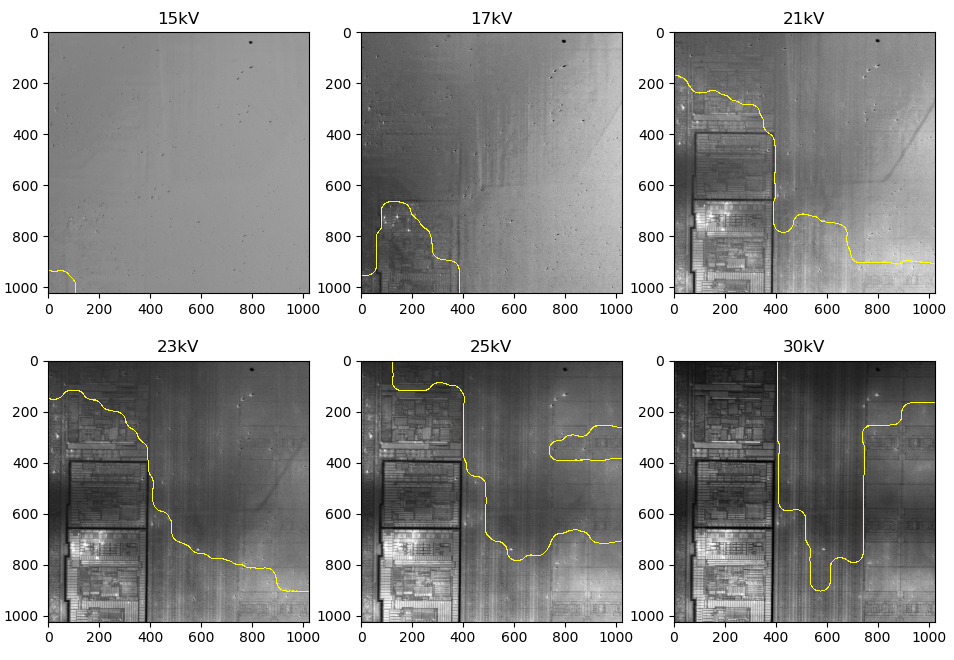}
    \caption{Contour maps of BSE data taken at different E-beam voltages at Location 1.}
    \label{fig: bse contours}
\end{figure}

\begin{figure}
    \centering
    \includegraphics[width=.75\linewidth]{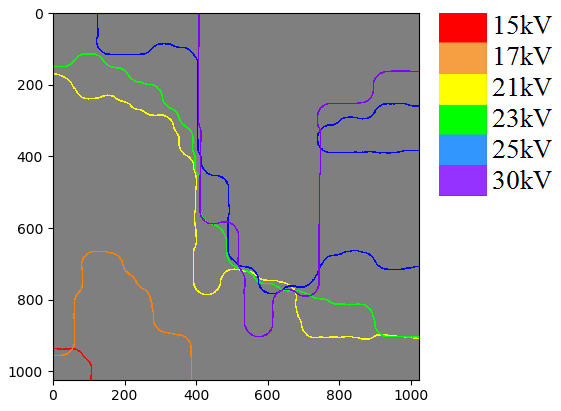}
    \caption{All contour maps of BSE data taken at various E-beam voltages at Location 1, overlayed.}
    \label{fig: bse all contours}
\end{figure}

The extracted contours for location 1 (from Fig.~\ref{fig: image collect}) are shown in Fig.~\ref{fig: bse contours} for each accelerating voltage, plotted on the corresponding images. The contours separate the boundary between regions with high versus low feature visability. However, there are contouring challenges when the intensity changes drastically. For example, between the images taken at 21~kV and 23~kV, the brightness distribution is much darker overall. Thus, when plotting all contours on the same plot, as in Fig. \ref{fig: bse all contours}, the contours at 21~kV and 23~kV are very close, whereas the contours at other voltages show more of a spreading trend as the voltage increases.   

\subsection{Estimated Thickness Measurement using Monte Carlo Simulation}
\begin{figure}
    \centering
    \includegraphics[width=1\linewidth]{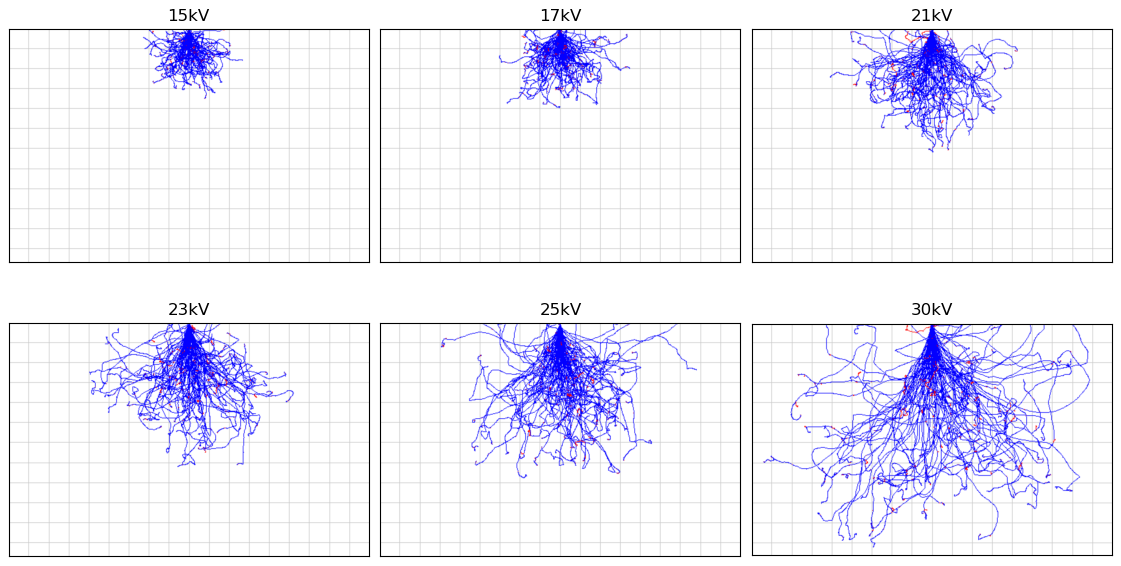}
    \caption{Monte Carlo simulation of electron trajectories in a bulk sample of silicon at 15, 17, 21, 23, 25, and 30~kV. Each division is 500~nm by 500~nm. Simulations were obtained using EISS software~\cite{eiss}. Trajectories of backscattered electrons are in red while trajectories of inelastically scattered electrons are in blue.}
    \label{fig: results_monte_carlo}
\end{figure}

Monte Carlo simulations of the electron-silicon interactions conducted at the the six accelerating voltages as mentioned in Sec.~\ref{Sec:Sample_prep} (E-beam voltage at 15, 17, 21, 23, 25, and 30~kV) and corresponding electron trajectories in bulk silicon presented in  Fig.~\ref{fig: results_monte_carlo}. 
As the E-beam energy is increased, the interaction volume increases (see Fig.~\ref{fig: results_monte_carlo}). The penetration depth is the maximum depth where backscattered electrons can reach, discounting outliers. For example, at 15~kV, the penetration depth is approximately 1~$\mu$m while at 30~kV the penetration depth at is approximately 4~$\mu$m. This is consistent with the increased visibility of sub-surface features as the accelerating voltage is increased, which is demonstrated in Fig.~\ref{fig: bse contours}. 

\subsection{Validation using a Commercial Instrument}
\begin{figure}
    \centering
    \includegraphics[width=0.75\linewidth]{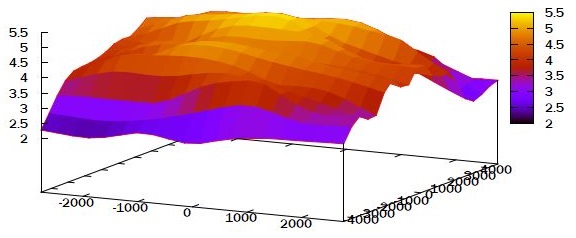}
    \caption{Real thickness map of the Xilinx FPGA after polishing. All axis values are in $\mu$m.}
    \label{fig: results_thickness_map}
\end{figure}

The real thickness map of the Xilinx FPGA test sample obtained from a commercial instrument is shown in Fig. \ref{fig: results_thickness_map}. The average thickness of the lower left quadrant of Location 1 is 2.2 $\mu$m. From the contour maps in Fig.~\ref{fig: bse contours}, the accelerating voltage at which the selected section is mostly visible is about 21~kV. Then, from the Monte Carlo simulations in Fig.~\ref{fig: results_monte_carlo}, the penetration depth at that voltage is approximately 2~$\mu$m. This depth is consistent with the real thickness map. 

\section{Conclusion and Future Work}
The real thickness of an ultra-thinned die is nonuniform. In this paper, superpixel segmentation was used to gain insight on the visibility of die structures at different accelerating voltages. These insights were used in combination with Monte Carlo simulation to roughly estimate the thickness of the remaining silicon. 

To improve segmentation accuracy, future work will involve obtaining multiple images and averaging them instead of using only one image. In addition, the precision of the thickness measurement estimation can be increased by reducing the step size between successive accelerating voltages. Though the preliminary results are not yet ideal, they are promising as a first step. 

Our next endeavor would be to use the contour maps and Monte Carlo simulation to predict the thickness of a whole die silicon. The predicted thickness map will then be used as online feedback to adjust FIB parameters for automated, uniform delayering for physical assurance and inspection of electronics.

\section*{References}

\def\refname{}


\end{document}